\documentclass{article}
\usepackage{graphicx, amsfonts, authblk, amsmath}
\newcommand\blfootnote[1]
{
  \begingroup
  \renewcommand\thefootnote{}\footnote{#1}
  \addtocounter{footnote}{-1}
  \endgroup
}
\newtheorem{definition}{Definition}

\usepackage[framemethod=tikz]{mdframed}
\usepackage[a4paper, margin=1.75in]{geometry}

\surroundwithmdframed[
  linewidth=0.5pt,
  roundcorner=4pt,
  backgroundcolor=gray!5,
  linecolor=gray!40,
  innertopmargin=6pt,
  innerbottommargin=6pt,
  innerleftmargin=10pt,
  innerrightmargin=10pt
]{definition}

\title{Decentralised Multi-Manager Fund Framework: Operationalising AutoVaults (v0.1)}
\author{Arman Abgaryan*$^{I}$, Utkarsh Sharma*$^{I, II}$, Joshua Tobkin$^I$}
\date{January 2025}

\begin{document}

\maketitle

\blfootnote{*Lead co-authors and designers.}
\blfootnote{I. Supra DeFi Research}
\blfootnote{II. Dept. of Engineering Science, University of Oxford}

\begin{abstract}
    We introduce a general-purpose, fully decentralised and algorithmic framework for permissionless multi-strategy capital allocation, implemented through tokenised automated vaults (AutoVaults) that consume signals from canonical signal oracles and leverage decentralised automation services for generation of execution directives. The framework is designed to function analogously to a multi-strategy asset management system, which is implemented entirely on-chain through a modular architecture comprising four interacting layers. The first is a capitalisation layer, which consists of vault smart contracts that accept multi-asset deposits, tokenise depositor participation, and enforce high-level risk limits and admissible venues for capital deployment. The second is the universal financial controller layer \cite{abgaryan2024intralayer}, which includes strategy contracts and Canonical Signal Oracles (CSO) states, in which the strategy contracts ingest CSO states and other data to synthesise execution directives governing the vault, thereby provisioning a decentralised marketplace governed by an adversarial and gamified validation mechanism. The third is an execution layer, which operationalises strategies using the host blockchain network's services, with a specific focus on automation. The fourth is a validation layer, which may be used to validate strategies and CSO states, enabling capital reallocation towards those exhibiting superior risk-adjusted performance. In this design, each admitted strategy acts as a manager for the \lq\lq fund\rq\rq, controlling assets deposited in smart contract vaults that issue transferable \textit{V-Tokens}, which represent fractional ownership of the real-time portfolio. The system is open to both human and AI agents alike, each of whom may serve as capital allocators, vault deployers, strategy developers, CSOs, or validators. The structure resulting from a quartet of these layers is a self-regulating asset management ecosystem capable of decentralised, cooperative, end-to-end optimisation across traditional and digital financial domains. This framework requires a host chain network with native automation and oracle services, enabling strategies to operate autonomously on-chain and paving the way toward self-sufficient, dynamically allocated capital across diverse objectives.
\end{abstract}

\section{Introduction}
    In traditional finance, multi-strategy and/or multi-manager companies have become a compelling structure for active (or passive) management of investor funds\footnote{Some multi-strategy firms integrate market making desks alongside investment management strategies, allowing them to diversify capital deployment across trading styles, and have diversified streams of trade execution.}, involving dynamic allocation across a wide range of validated strategies, whilst enabling specialised teams to operate within their individual\lq\lq pod\rq\rq. The proposed framework emulates a multi-manager architecture for decentralised finance, offering a superior method of trust-minimised capital management which has thus far been dominated by centralised, single-strategy models, with rigid capital silos, and inefficient or opaque validation and allocation mechanisms.\\
    \\
    The Decentralised Multi-Manager Fund (DMMF) framework, seeks to provide an entirely decentralised, permissionless, and modular architecture that transforms decentralised capital management from static, singular strategies into a competitive and adaptive marketplace, where diverse strategies are continuously assessed and optimally allocate capital. Notably, whilst capital allocation is primarily focused at profit maximisation, allocation may also be focused on broader objectives.\\
    \\
    The DMMF framework achieves the following objectives:
    
    \begin{itemize}
        \item \textit{Trust-minimised execution}. Capital management, both on-chain and off-chain, may be vulnerable to significant failures due to limited safeguards and opaque execution practices, such as reliance on manual intervention for execution or unreliable automation. We address this challenge by embedding execution within a decentralised, automated infrastructure stack, leveraging the host blockchain network's trust-minimised Layer 1, natively-integrated automation and oracle services \cite{chakka2023dora}, bridging \cite{kate2025hyperloop}, and IntraLayer \cite{abgaryan2024intralayer} coordination, to ensure transparency and dependability. In this setup, vaults are designed with strict executional constraints and programmable parameters, defining ex-ante where and how capital can be deployed, thereby minimising discretionary control and trust assumptions by design.
        \item \textit{Strategy discovery at scale}. Current capital management ecosystems rely on a narrow set of insiders for strategy proposals, limiting competition, diversity, and adaptability to heterogeneous market conditions. We seek to maximise the search space for active and passive strategies through open, permissionless contribution and evaluation mechanisms, which could be achieved with creation of a marketplace of strategies and CSOs.
        \item \textit{Segregation between insight publishers and users}. The framework decouples signal production from its end consumers, i.e. agents with proprietary insights publish data by creating and maintaining Canonical Signal Oracles (CSO) states, while strategy contracts (acting as asset managers) select which of the available states can be utilised.
        \item \textit{Community-owned}. Currently, governance is fragmented and lacks a native layer for expressing ownership over fund performance. This framework can be applied to tokenise ownership, enabling liquid, composable, and programmable exposure to each vault's economic outcomes, and establish a collective management mechanism that allows owners to determine the strategic course of the vault.
        \item \textit{Dynamic allocation}. Often, capital managed through vaults operate under rigid allocation logic tied to a single strategy, lacking adaptability to changing market conditions. We seek to enable vaults to define the multi-strategy search space - specifying objectives, constraints, and risk tolerance, while meta-strategies acting as competing managers can optimise allocation to multiple strategies within that space through continuous, performance-sensitive logic. This allocation can be further supported by a crowd assessment framework for strategy validation, which helps ensure that the permissionless and decentralised nature of the proposed framework does not erode performance.
        \item \textit{Best execution}: Asset management frameworks require \lq\lq best\rq\rq execution of orders, which can be achieved within the framework through IntraLayer \cite{abgaryan2024intralayer}, which aggregates liquidity as its Universal Financial Controller (UFC)'s are able to route orders to vaults deployed across multiple blockchains and qualified off-chain systems.
    \end{itemize}

    \noindent
    The proposed framework reimagines decentralised capital management as a \lq\lq pod shop\rq\rq multi-manager ecosystem, where a quartet of interlinked and synergetic layers, unlocks the potential for a competitively self-sustaining capital management ecosystem. These layers are:

    \begin{itemize}
        \item \textit{Autovault-based Capitalisation layer}: The layer incorporates intention-gated vaults that accept multi-asset deposits, and mint fungible vault-share tokens. These vaults specify which \lq\lq capital-manager\rq\rq strategy contracts (allocators) may deploy the capital downstream, providing a clean separation between ownership, specifying constraints within which assets are to be managed, and execution.
        \item \textit{UFC layer}: A permissionless marketplace of smart contract–based strategies and proprietary CSO state feeds, where strategy contracts process CSO states and on-chain data to generate automated execution orders for managing assets submitted to the vaults, which may include orders to multiple blockchains.
        \item \textit{Execution layer}: Using the host blockchain network's oracle, block-level automation\footnote{Block-level automation refers to a native, transaction engine embedded in the block construction. It automatically triggers submitted smart-contract transactions, eliminating the need for external bots or re-layers.} and efficient cross-chain execution (IntraLayer infrastructure), transactions can be efficiently (and autonomously) executed, allowing strategy contracts to fulfil their mandates.
        \item \textit{Validation \& Allocation layer}: Each proposed strategy and CSO states consumed by the strategy contract, may enter an adversarial validation game (for e.g., OpenAlpha \cite{abgaryan_openalpha_2025}), undergoing predictive scoring, stake weighted challenges, audits, and prediction market voting to prove compliance with the originating intention, such that strategies that clear this gauntlet become eligible for allocation. Once a subset of validated strategies is established, they may function as capital management agents within the vault as a single vault may have multiple managers, where algorithmic allocators continuously rebalance weights across an active set of strategies according to rules pre-programmed (or voted on) in the vault contracts or specialised meta-strategies responsible for rebalancing. Here, capital is redirected toward strategies exhibiting persistently superior risk-adjusted returns, thereby diversifying exposures and mitigating systemic concentration risk.
    \end{itemize}

    \noindent
    At its core, the framework encapsulates multiple strategies acting as managers within smart contract vaults, each vault issuing tokenised fractional ownership units, termed \textit{V-Tokens}, representing transparent, real-time exposure to actively managed strategy portfolios, and also providing proportional governance rights over the management of assets in the vault (e.g., rebalancing, how the assets are deployed, control of the assets, etc.). Each vault is defined by an explicit parametric tuple: $V = (\mathbf{A}, \mathbf{S}, \mathbf{O})$, where $\mathbf{A}$ represents the set of permissible assets, $\mathbf{S}$ represents the vector of tuples defining the strategy IDs and the corresponding amounts (share in the fund) that can be managed by a particular strategy; $\mathbf{O}$ is a vector of tuples specifying the venues along with the list of functions or operations applicable for managing capital in each context.

    \begin{definition}[Net Asset Value]
        The Net Asset Value (NAV) of a vault at time $t$ is defined as the liquidation value of the vault's current holdings, net of accrued fees:
        \begin{equation}
            NAV^{V}_t = \sum_{i=1}^{m} q_{i,t} \cdot P_{i,t} - f^{V}_t,
        \end{equation}

        \noindent
        where $q_{i,t}$ denotes the quantity of asset and positions $i$ held at time $t$, $P_{i,t}$ the corresponding market price, and $f^{V}_t$ cumulative fees owed by the vault\footnote{We assume that fees are not already deducted from the vault's cash ledger as they accrue.}.
    \end{definition}

    \noindent
    Note, that the framework's design implicitly fosters a marketplace where human or algorithmically autonomous (AI-driven) participants symmetrically compete and cooperate. This is achieved by harmonising incentives among fund managers, strategy proposers, CSOs, validators, and execution agents, which ensures competitive expedience for reallocation of capitals across strategies, incentivised pressure to innovate compelling strategies and proprietary signals, and these to be provided with best execution.

\section{Architecture}
    Before describing Decentralised Multi-Manager Framework's architecture, we state definitions for clarity:
    
    \begin{definition}[Canonical Signal Oracles]
        A Canonical Signal Oracle (CSO) is an agent ($i$) identified by unique account addresses and maintain vector of state tuples $\mathbf{\Sigma}^i = (\mathbf{\kappa}, \mathbf{D})$, where $\kappa \in \mathcal{K}$ specifies the type of capital management activity (with $\mathcal{K}$ the set of all such activities), and $\mathbf{D}$ is a vector of state data associated with the CSO.        
    \end{definition}
    
    \noindent    
    The CSO's state space is the disjoint union spanning $\bigcup_{\kappa \in \mathcal{K}} \{\kappa\} \times \mathbf{D}^\kappa$, with $\mathbf{D}^\kappa$ representing the domain of admissible state data for the type of capital management activity $\kappa$. The disjoint-union structure enforces type separation across management roles, while CSO states tied to different activities may share a common semantic representation through their respective fields. This design enables strategies and validators to query, compare, and compose them without ambiguity.\\
    \\
    Some examples of such CSOs are:
    
    \begin{itemize}
        \item Portfolio allocation CSOs, which may publish a portfolio weight vector $\mathbf{w}_t \in \mathbb{R}^n$ or a categorical decision vector $\mathbf{c}_t \in \{-1,0,1\}^n$ indicating buy, hold, or sell actions. 
        \item Market-making CSOs, which may expose bid–ask quotes with volumes $\{(b_{i,t}, v^b_{i,t}, a_{i,t}, v^a_{i,t})\}_{i=1}^n$, thereby encoding snapshots or updates of a limit order book.
        \item Arbitrage CSOs, which may publish tuples of trade legs with implied spreads or mispricings, $\{(\ell^1_t, \ell^2_t, \dots, \ell^k_t, \Delta_t)\}$, specifying coordinated opportunities across venues.
        \item Liquidity-provisioning CSOs, which may publish allocations $\{(p_j, a_j, d_j)\}_{j=1}^m$, where $p_j$ is a pool identifier, $a_j$ the allocated notional, and $d_j$ the lock-up duration. 
        \item Yield-oriented CSOs, which may provide tuples $\{(c_i, y_{i,t}, d_i)\}_{i=1}^k$ capturing instrument identifiers, quoted yields, and maturities of income-focused positions.
        \item Reward-allocation CSOs (e.g., liquidity mining), which may publish distribution parameters such as a Merkle root referencing a tree of rewards, ensuring on-chain verifiability, or parameter sets that define reward distribution logic (e.g., emission schedules, or coefficients related to matching rules).        
    \end{itemize}
    
    \begin{definition}[Strategy]
        A strategy is a parametric function that is expressed through a smart contract, such that it maps available external market data from decentralised oracle service\footnote{A decentralised oracle service is a distributed network that provides reliable external data to smart contracts without reliance on a single trusted source. By aggregating inputs from multiple independent nodes, it mitigates the risks of manipulation, ensuring secure and verifiable data delivery for on-chain applications.}, relevant state data natively generated on-chain, and CSO states to executable actions on vaults, using block-level or off-chain automation, subject to a pre-defined set of venues and asset universe, whilst potentially adhering to vault-specific constraints.
        
        \begin{equation}
            S: (\mathcal{M}_t, \mathcal{X}_t, \theta_S) \mapsto \mathcal{E}_t,
        \end{equation}

        \noindent
        where $\mathcal{M}_t$ is the market state (e.g., prices, liquidity, state data generated on-chain); $\mathcal{X}_t$ is the CSO space; $\theta_S$ represents the strategy-specific parameters; and $\mathcal{E}_t$ represents the executable actions (e.g., trade orders)
    \end{definition}

    \begin{definition}[Capital Vault]
        A capital vault ($V = (\mathbf{A}, \mathbf{S}, \mathbf{O})$) can be defined parametrically using a tuple, comprised of the admissible asset set ($\mathbf{A}$), admitted strategy set that can manage those assets - with allocation shares $\alpha_j$ such that $\mathbf{S} = \{(S_j, \alpha_j)\}_{j=1}^{k}$, and execution environment ($\mathbf{O} = \{(v_i, \mathcal{F}_i)\}_{i=1}^{m}$) capturing allowed venues $v_i$ and permitted operations $\mathcal{F}_i$. It enforces capital constraints and controls execution, through on-chain programmable rules, governance, interface with trading venues and wider DeFi protocols.
    \end{definition}

    \begin{definition}[Validation Mechanism]
        A validation protocol is a mechanism that maps submitted strategies or signals to a binary accept/reject decision. For e.g., in the case of OpenAlpha \cite{abgaryan_openalpha_2025} - a multi-stage adversarial mechanism, using prediction markets and incentivised agents to reach probabilistically robust outcomes.
    \end{definition}

    \noindent
    In the framework, capital allocation is intention-gated and competitively optimised, such that each vault captures capital owners' intentions $\mathcal{I}_V$, which encodes parametric objectives over performance metrics (e.g., stability, return, drawdown). This compels strategy proposers to design candidate strategies, while CSOs publish signals that can be consumed by different strategies, where transactions representing function calls in strategy contracts may be triggered by the host network's automation service (e.g., block-level automation). These strategies are submitted to the marketplace, which may be linked to specific capital vaults, and community validation based mechanism (e.g., OpenAlpha \cite{abgaryan_openalpha_2025}) can be used to qualify strategies to be used as \lq\lq manager\rq\rq in vault. Together, the strategy contract and CSO function as Universal Financial Controllers \cite{abgaryan2024intralayer}. In the context of cross-chain execution, the UFCs move assets and execute across multiple blockchains, and the capital vaults may operate as IntraLayer vaults, that use the cross-chain bridge and system's liquidity network. These controllers autonomously route funds and instructions across chains and systems, whereby automation triggers conditional execution(\cite{abgaryan2024intralayer}).\\
    \\
    At the lowest level, IntraLayer framework (with decentralisation automation) continuously bridges liquidity across heterogeneous blockchains, and routing orders to the venue for best execution (through DFMM), thereby allowing a single vault to aggregate fragmented liquidity and managed by multi-chain, multi-system strategies in a trust-minimised manner. Altogether, this provisions the vaults into fully digital, multi-manager system provisioning potentially open-sourced asset management, where strategy logic, risk controls, and execution quality are enforced by smart contracts, and capital can flow seamlessly between decentralised markets without custodial friction.\\
    \\
    The proposed framework's architecture can be broadly described with the following key components:

    \begin{enumerate}
        \item Deployment: The vault deployer deploys the vault contract on the host blockchain network, which is comprised of six core modules:
            \begin{itemize}
                \item \textit{Venue Registry}, lists external marketplaces, applications, and corresponding functions where the vault would deploy its assets.
                \item \textit{Strategy Registry}, stores IDs and permissions for whitelisted strategies (strategy owners).
                \item \textit{Share Tokenisation}, mints fungible V-tokens against deposited collateral.
                \item \textit{Governance Module}, V-Token holders govern whitelists, parameters, and other aspects of the vault.
                \item \textit{Accounting Module}, records NAV and performance attribution at vault and strategy level.
            \end{itemize}        
        \item Capitalisation: Investors deposit assets, and the contract issues V-tokens representing pro-rata exposure and governance rights.Governance whitelists execution venues (on a single chain or across multiple chains) and sets capital limits per venue.  
        \item  CSO On-boarding: The signal marketplace contract lets CSOs register new signal objects on-chain, which can be consumed by strategies upon subscription.
        \item Strategy On-boarding: Strategy proposers implement the smart contract logic of their strategy and deploy it on the host blockchain network, and request listing in the Strategy Registry so it can interact with the vault.This paves the way for a marketplace of strategies to emerge, as candidates for whitelisting (through governance), without prohibitively lofty barriers to entry. Strategies can be managed either by appointed managers, through a decentralised governance mechanism, or both.
        \item Community Validation and Dynamic Allocation: A decentralised layer (for e.g., OpenAlpha \cite{abgaryan_openalpha_2025}) where stakers audit strategies, signals (when possible) and end-applications, score them against the vault’s intent specification, and drive stake-weighted capital routing and real-time rebalancing. Based on community validation or other decision criteria, the vault’s governance may approve strategies as capital managers. Note that whitelisted strategies cannot arbitrarily move vault's assets, as they only submit execution intents, which are constrained by vault rules. 
         \item The decentralised automation service triggers transactions in the strategy contract, which—per its logic consumes CSO states, decentralised-oracle feeds, and on-chain data to derive executable capital-management decisions, these decisions invoke functions on the vault contracts, either directly or via a cross-chain bridge   
         \item When invoked by the strategy contract, the vault contract interfaces with DeFi applications to deploy and manage the vault’s capital.        
    \end{enumerate}

    \noindent
    This framework's architecture can be visualised in the summary schematic, as follows:
    
    \begin{figure}
        \begin{center}
            \includegraphics[scale=0.40]{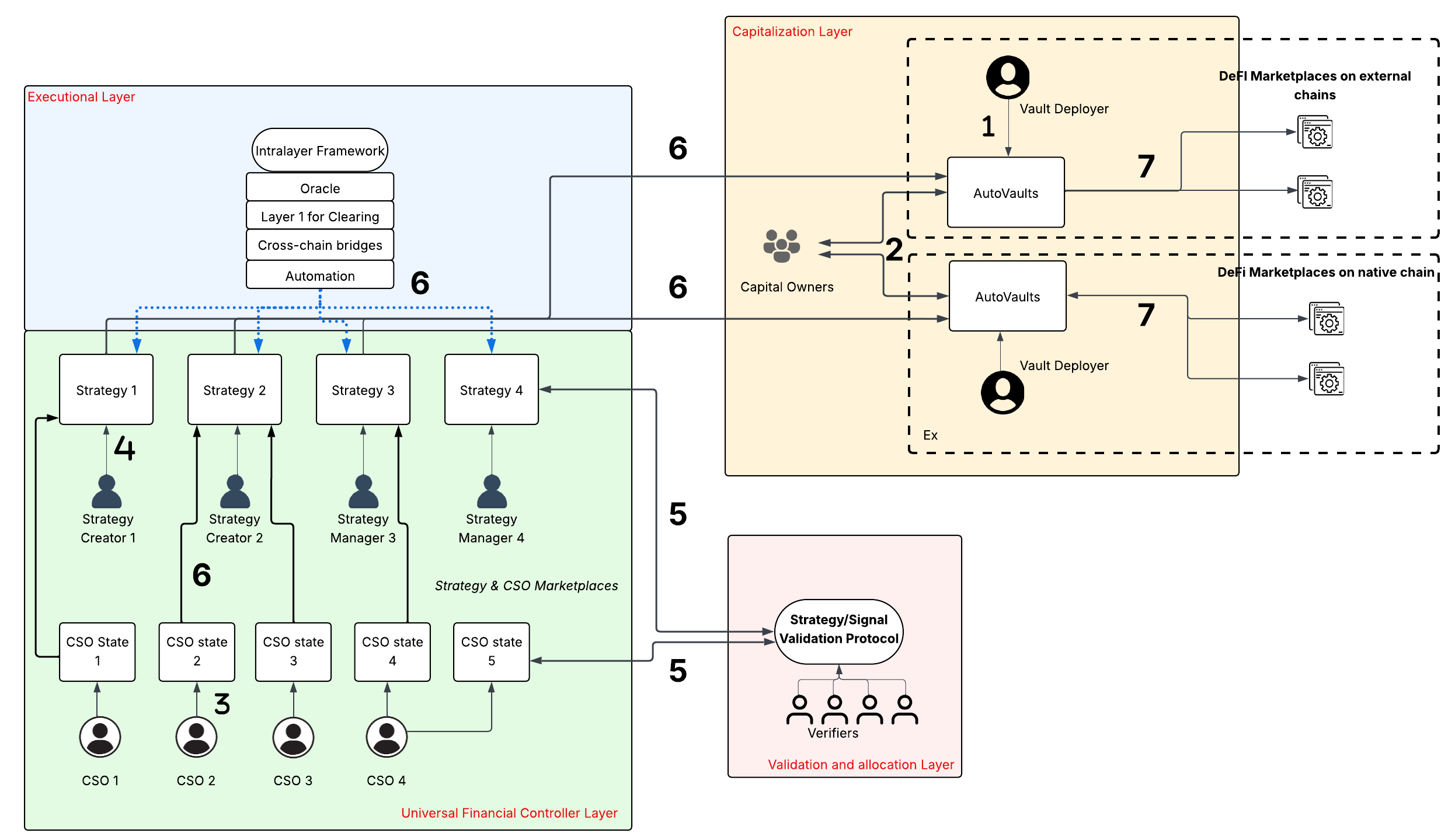}            
            \caption{DMMF Summary schematic.}
            \label{fig:blocks}
        \end{center}
    \end{figure}

\section{Signals Marketplace}
    Signals can be viewed as CSO state updates, that are consumed by strategy contracts, functioning analogously to data oracles, but generalised to encode any asset-management directive. The proposed framework enables contributors to keep proprietary strategies (for e.g., custom market making algorithms, trading strategies, etc.) private by updating only the state, such as allocation signals, in the form of weight vectors to target assets, rather than disclosing full executable strategy code on-chain. This incentivises researchers to develop their strategies, signals, and continuously enhance their approach, without fear of disclosing their proprietary methods. Instead of disclosing full algorithmic logic, signal providers conduct periodic state updates, where the state can represent various useful forms of data, such as allocation vectors ($\mathbf{w}_t = [w_{1,t}, w_{2,t}, \dots, w_{n,t}]$) indicating recommended weights for $n$ assets. These signals are a universal abstraction for representing any portfolio action or financial activity, as different signal types can encode:
    \begin{itemize}
        \item \textit{Allocation vectors}: portfolio weights or categorical buy/hold/sell indicators over assets.
        \item \textit{Arbitrage bundles}: tuples of coordinated trade legs with implied spreads across venues.
        \item \textit{Market-making quotes}: bid/ask prices with corresponding volumes, effectively encoding a limit order book snapshot.
        \item \textit{Liquidity provisioning instructions}: pool identifiers, notional allocations, and lock-up parameters.
        \item \textit{Yield/funding offers}: instrument IDs with quoted yields and maturities.
        \item \textit{Allocator distributions}: rules for conditional fund routing, milestone-based disbursement, or matching contributions.
    \end{itemize}

    \noindent
    In practice, signals contained in CSO states unify diverse financial actions, ranging from portfolio reallocations to market-making quotes. Smart contracts, triggered by the host chain’s native automation layer, consume these signals and translate them into actions within the vaults, while off-chain models and datasets remain private. For example, these can include allocation vectors or categorical buy/hold/sell indicators capture standard portfolio adjustments, while technical indicators such as moving-average crossovers or volatility-regime flags extend to dynamic trading styles. Similarly, sentiment-based updates derived from news or social media can be encoded as state changes, and market-making activity can be represented through bid/ask quotes with associated volumes. This design cleanly separates off-chain signal generation, where proprietary datasets and models remain private, from on-chain execution, where only final outputs are published and enforced by smart contracts.\\
    \\
    Strategies may subscribe to read allocation CSO states under a defined fee structure, with two primary models:
    \begin{itemize}
        \item \textit{Subscription model}: CSO earn a fee for every new strategy using their signal, a structure well suited to passive or beta-style contributions. 
        \item \textit{Participation model}: CSO commit their own capital alongside their signals, aligning incentives and allowing them to capture a share of realised returns in addition to any baseline fee. This model encourages active researchers to develop and broadcast higher-quality signals.  
    \end{itemize}

    \noindent
    The strategy contracts may implement logic that aggregates subscribed signals into portfolio-level directives, with weights determined by performance, or governance parameters, and rebalances as new signals arrive. For example, if $k$ providers each submit a weight vector $\mathbf{w}^j_t = [w^j_{1,t}, w^j_{2,t}, \dots, w^j_{n,t}]$ for $n$ assets at time $t$, the aggregate capital managed by strategies subscribed to the signal, may act as an aggregator of heterogeneous signals, which may be weighted by the total capital of subscribed strategies and their realised performance,

    \begin{equation}
        \mathbf{W}_t = \frac{1}{Z} \sum_{j=1}^{k} s_j \mathcal{P}_j g_j \mathbf{w}^j_t,
    \end{equation}
    
    \noindent
    where $Z = \sum_{j=1}^{k} s_j \mathcal{P}_j g_j$ is the normalising factor, $s_j$ denotes the capital managed by subscribed strategies $j$, and $\mathcal{P}_j$ is the performance weight, such that to enforce risk discipline, allocations are constrained by $\sum_{i=1}^{n} |W_{i,t}| \leq 1$, with $|W_{i,t}| \leq \bar{w}$, where $\bar{w}$ is a governance-set per-asset cap.\\
    \\
    This can be used to quantify fees, and external insurance protocols could emerge around this system to underwrite strategy performance or to offer protection against poor signal reliability. Altogether, the proposed mechanism may aid in mitigating problems that arise when back-tested strategies perform poorly out of sample, albeit at the cost of subscribers who pay on the basis of \textit{expected} rather than \textit{realised} returns.\\
    \\
    In addition to deterministic fee allocation, the framework can accommodate auction-based mechanisms in which subscribers bid for access to high-value signals which can create a competitive marketplace for signal provision, where access prices are endogenously determined by demand. The resulting fee flows operate as a form of revenue sharing, directly funding the production of fundamental research and the refinement of predictive signals. By aligning compensation with demonstrated market value, this approach promotes continuous innovation and improved price discovery, in contrast to traditional asset management models that rely on fixed management and performance fees largely detached from signal quality.\\
    \\
    This approach enhances the strategy layer by allowing contributors to protect proprietary models and data, broadening participation to anyone capable of generating high-quality signals (CSO state updates), and creating modular revenue streams through flexible subscription or profit-sharing mechanisms. It enables continuous, permissionless onboarding of CSOs without disclosing sensitive intellectual property. By combining signals with permissionless smart contract execution, vault-level composability, dynamic signal pricing, open community validation, and cross-chain liquidity routing, the system becomes a universal marketplace for predictive allocation vectors. Signals can be reused across multiple strategy smart contracts, blended with fully disclosed on-chain strategies, competitively priced via auctions or subscription. The framework further extends quantitative allocation performance-weighted signal mixing, on-chain assessment of signals, and dynamic multi-vault routing aligned to diverse intent vectors. The combined effect is a dynamic, self-learning, fully composable signal ecosystem that continuously adapts allocation weights to realised risk-adjusted performance, while maintaining contributor privacy and verifiable trustlessness.

\section{Strategy Instantiation}
    DMMF defines a modular and parametric architecture for tokenised, automated   strategies and vaults capable of interchain (and cross-chain) execution of diverse investment logics, which, rather than a monolithic product, the system can be interpreted as a parametric family of strategy generators, each specified by a set of parameters: (i) \textit{asset universe}, i.e., the admissible set of tokens and venues; (ii) \textit{strategy class}, defining the formal investment or trading algorithm; (iii) \textit{execution frequency}, determining cadence for rebalancing, contract rollovers, or dynamic adjustments; (iv) \textit{risk envelope}, specifying leverage, drawdown, and exposure constraints; and (v) \textit{execution logic}, encoding order generation and trade initiation. These parameters provide levers for deploying and customising a wide range of strategy variants with minimal additional audit overhead.\\
    \\
    Within this framework, strategy contracts may include:
    
    \begin{enumerate}
        \item \textit{Spot strategies}: Direct, unlevered exposure to digital assets, emphasising capital efficiency and low-friction execution.
            \begin{itemize}
                \item \textit{Pure spot}: Vaults hold positions in assets such as ETH, BTC, or stablecoins, for use cases including swaps and reserve management.
                \item \textit{Staked spot}: Assets are allocated to staking or restaking protocols to generate yield while maintaining base-asset exposure.
            \end{itemize}
        \item \textit{Linear-derivative strategies}: Deployment of capital through linear instruments (futures, forwards), with on-chain mechanisms for roll-over, collateralisation, and margin management.
            \begin{itemize}
                \item \textit{Index}: Tracking of broad-market or thematic baskets, rebalanced periodically to minimise tracking error.
                \item \textit{Basis capture}: Exploiting funding rate differentials (e.g., spot–perpetual arbitrage), with dynamic monitoring and unwind thresholds.
            \end{itemize}
        \item \textit{Non-linear derivative strategies}: Provision of asymmetric exposures using option-like or path-dependent instruments, aimed at risk-seeking participants. 
        \item \textit{Active strategies}: Discretionary or rules-based strategies requiring dynamic monitoring and allocation.
            \begin{itemize}
                \item \textit{Arbitrage}: Exploitation of cross-market or cross-chain inefficiencies, optimised for execution quality across fragmented liquidity. 
                \item \textit{Liquidity management}: Reallocation of liquidity across DeFi venues, providing LP exposure under dynamic conditions.
                \item \textit{Market making}: Continuous posting of bid/ask quotes across venues and instruments, managing spreads and inventory risk.
            \end{itemize}

        \item \textit{Thematic strategies}: Vaults that are directed toward non-commercial or collective objectives.
            \begin{itemize}
                \item \textit{Grant distribution}: Milestone-based funding disbursements conditional on predefined criteria.  
                \item \textit{Liquidity mining}: Non-profit deployment to bootstrap liquidity in emerging protocols.  
                \item \textit{Airdrop strategies}: Token distribution to incentivise governance or community participation.  
                \item \textit{Matching funds vaults}: Capital matching mechanisms (e.g., quadratic funding) to enhance fairness in public-goods finance.  
                \item \textit{Bug bounty vaults}: Allocation of rewards for responsible vulnerability disclosure.  
                \item \textit{Sustainability and impact vaults}: Capital directed toward environmental or social objectives (e.g., carbon offsets, humanitarian aid).
                \item \textit{Protocol insurance funds}: Reserve accumulation to insure against smart-contract failures or systemic risks.  
            \end{itemize}
    \end{enumerate}

    \noindent
    In essence, all strategies and associated vaults rely on the same set of underlying layers (allocation, signal proposition, validation, and execution).

\section{Conclusion}
    We introduced a conceptual multi-layered framework for Canonical Signal Oracles (CSOs) and Universal Financial Controllers (UFCs) to enable modular, verifiable, and capital-efficient composition of on-chain strategies, atop automated vaults. The proposed architecture is organised into four interdependent layers: (i) a capitalisation layer, which tokenises deposits into vault shares and enforces admissible risk and venue constraints; (ii) a UFC layer, where strategy contracts consume CSO states to generate execution directives; (iii) an execution layer, which operationalises these directives through decentralised automation and cross-chain infrastructure; and (iv) a validation \& allocation layer, which continuously evaluates signals and strategies to redirect capital toward those with superior risk-adjusted performance. Together, these layers establish a transparent and competitive multi-manager ecosystem, aligning incentives among vault participants, strategy developers, and validators. Altogether, the work presents a composable, verifiable financial stack in which capital, strategies, and signals interact through the proposed framework, potentially enabling more transparent and resilient markets.\\
    \\
    In future iterations, we will formalise specifications, and provide relevant proofs, supplemented with a comprehensive literature review.
    
\bibliography{main}

\begin{thebibliography}{1}

\bibitem{abgaryan2024intralayer}
Arman Abgaryan and Utkarsh Sharma.
\newblock Intralayer: A platform of digital finance platforms.
\newblock {\em arXiv preprint arXiv:2412.07348}, 2024.

\bibitem{abgaryan_openalpha_2025}
Arman Abgaryan and Utkarsh Sharma.
\newblock Openalpha: A community-led adversarial strategy validation mechanism for decentralised capital management.
\newblock {\em arXiv:2506.21809 [q-fin.GN]}, June~13 2025.
\newblock arXiv preprint arXiv:2506.21809.

\bibitem{chakka2023dora}
Prasanth Chakka, Saurabh Joshi, Aniket Kate, Joshua Tobkin, and David Yang.
\newblock Dora: Distributed oracle agreement with simple majority.
\newblock {\em arXiv preprint arXiv:2305.03903}, 2023.

\bibitem{kate2025hyperloop}
Aniket Kate, Easwar~Vivek Mangipudi, Charan Nomula, Raghavendra Ramesh, Athina Terzoglou, and Joshua Tobkin.
\newblock Hyperloop: Rationally secure efficient cross-chain bridge.
\newblock {\em Cryptology ePrint Archive}, 2025.

\end{thebibliography}
\bibliographystyle{plain}

\end{document}